\documentclass[twoside,twocolumn,english,aps,prb]{revtex4-2}

\usepackage[usenames,dvipsnames]{xcolor}
\usepackage{amsmath}
\usepackage{amsfonts}
\usepackage{amssymb}
\usepackage{stmaryrd} 
\usepackage[T1]{fontenc}
\usepackage{mathptmx}
\usepackage[colorlinks=true,citecolor=blue,linkcolor=red]{hyperref}
\usepackage{graphicx}
\usepackage{bbold}	
\usepackage[makeroom]{cancel}		
\usepackage{multirow}				
\usepackage[normalem]{ulem}        
\usepackage{array}
\usepackage{comment}
\usepackage{pdfpages}

\newcommand{\unit}[1]{\,\mathrm{#1}} 
\newcommand{\equa}[1]{Eq.~\eqref{#1}} 
\newcommand{\fig}[1]{Fig.~\ref{#1}} 
 
\newcommand{\red}[1]{#1} 

\newcommand{\append}[1]{Appendix~\ref{#1}}
\usepackage{ulem}

\newcommand{\rom}[1]{\uppercase\expandafter{\romannumeral #1\relax}}

\graphicspath{{figures/}}

\makeatletter
\AtBeginDocument{\let\LS@rot\@undefined}
\makeatother

\begin{document}
\title{Supercurrent \red{growth in nonequilibrium superconductors}}
	
	\author{Qinghong Yang$^{1,\ast}$}
	\author{Yuqi Cao$^{1,\ast}$}
	\author{Dante M. Kennes$^{\red{2,3}}$}
	\author{Zhiyuan Sun$^{\red{1,4},\dagger}$}

	\affiliation{$^{1}$State Key Laboratory of Low-Dimensional Quantum Physics and Department of Physics, Tsinghua University, Beijing 100084, People's Republic of China\\
    \red{$^{2}$Institut f\"ur Theorie der Statistischen Physik, RWTH Aachen University, 52056 Aachen, Germany}\\
    $^{3}$Max Planck Institute for the Structure and Dynamics of Matter, Center for Free-Electron Laser Science (CFEL), Luruper Chaussee 149, 22761 Hamburg, Germany\\
    \red{$^{4}$Frontier Science Center for Quantum Information, Beijing 100084, People's Republic of China
    }
}

	\date{\today}

	\begin{abstract}
		In ultrafast experiments on superconductors, a pump laser pulse often heats up the electronic system and suppresses the density of superfluid electrons.	
		Subsequently, the electrons undergo a  cooling process \red{because of} electron-phonon thermalization so that the superfluid density recovers in time.
		We study the nonequilibrium electromagnetic response of the system in this cooling process.
		We show that if a supercurrent is initiated by a probe electric field pulse, an intriguing  phenomenon of `supercurrent growth' occurs,
		meaning that the net current grows in time with the increasing superfluid density. 
		Using the Boltzmann kinetic equation, we uncover its microscopic origin as the momentum-relaxing scattering of Bogoliubov \red{quasiparticle}s by impurities and phonons, in stark contrast to the widely accepted intuition that impurities always attenuate currents. 
		We further show that supercurrent growth has important experimental manifestations, including the ultrafast Meissner effect and an optical reflectivity exceeding unity.
	\end{abstract}

	\maketitle


\section{Introduction}	
Nonequilibrium superconductivity has been a topic of persistent research 
	since the 1960s~\cite{
		chang_nonequilibrium_1978,
		Elesin.1981,
		Kopnin.2001_book,
		Tinkham, 
		Larkin.2005_book, 
		gray,  
		demsar}.
Interesting phenomena in this field range from the phonon-bottleneck effect in  \red{quasiparticle} recombination~\cite{Rothwarf.1967}, microwave-enhanced superconductivity~\cite{Wyatt.1966, Eliashberg.1970}, light manipulated pairing~\cite{claassen_universal_2019,  Yu.2021, gassner_light-induced_2024},
to light induced superconducting-like states whose exact physical nature is under \red{intense} debate~\cite{
Fausti2011,Nicoletti2014,mitrano16,Nicoletti2018,cantaluppi_pressure_2018, Suzuki2019, Budden2021, rowe23,Zhang2018,Zhang2018a,Cremin2019, Niwa.2019, Buzzi.2020, Buzzi.2021, 
Isoyama2021FeSeTeLightEnhance,
Nishida.2023, Dodge.2023,
Zhang.2024,  Kennes2017,Babadi2017,Sentef2017,Chiriaco2018,Wang.2018, Kaneko.2019, suntt, 
 chattopadhyay_metastable_2025,
 michael2025giantdynamicalparamagnetismdriven, Wang2025}.
	The defining feature of a superconductor is its electromagnetic (EM) response  that  contains information of not only the vanishing resistance and the Meissner effect, but also the light-matter hybridization~\cite{Sun.2020_collective, Gabriele_non-linear_2021, Sellati_Josephson_plasmon.2023, Sellati_Josephson_plasmon.2025,
	Gedik.2026, potts2026cooperplasmon}. 
	Therefore, the EM response is one of the key properties of nonequilibrium superconductors~\cite{kennes17,kenneser,Chou.2017, Grankin.2025}, which is typically  measured in the transient regime in ultrafast experiments~\cite{Fausti2011,Nicoletti2014,Orenstein.2015, mitrano16,Nicoletti2018,cantaluppi_pressure_2018, Budden2021, rowe23, Zhang2018,Zhang2018a,Cremin2019, Niwa.2019,  Nishida.2023, Dodge.2023, Zhang_2D_spectroscopy.2023, Zhang.2024,  Salvador_2D_spectroscopy.2024, taherian_probing_2025,ZJz2026}.

\begin{figure}[t]
		\centering    
		\includegraphics[width=0.8 \linewidth]{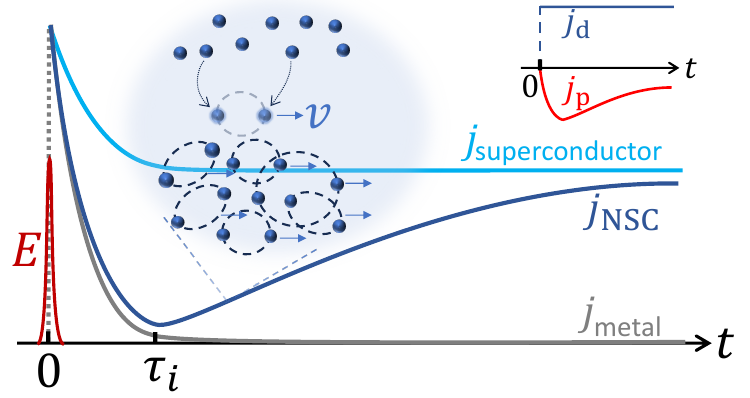}
		\caption{Illustration of the current in response to an electric field pulse (red curve) in time for different systems: normal metal (gray curve), superconductor (light blue curve), nonequilibrium superconductor (NSC) with a growing superfluid density (dark blue curve).
	  The left inset illustrates the microscopic process of the supercurrent growth in the NSC curve:  two electrons gain the  velocity $\mathbf{v}$ as they  form a Cooper pair. The top right inset schematically shows the time dependence of the diamagnetic current ($j_{\text{d}}$) and paramagnetic current ($j_{\text{p}}$) in the NSC current response.
		}
		\label{fig:1}
	\end{figure}

	The simplest approach to the EM response of superconductors, both in and out of equilibrium, is the time-dependent Ginzburg-Landau (TDGL) theory~\cite{Gorkov1968, Cyrot1973, Larkin.2005_book,frank_incompatibility_2016, kenneser}. The TDGL equation reads
	\begin{equation}\label{eq:tdgl}
		\frac{1}{\gamma}(\partial_t + i\phi)
		\psi 
		= \left[\alpha
		+\xi_0^2
		\left(
		\nabla-i \mathbf{A}
		\right)^2\right]
		\psi - 2 |\psi|^2 \psi
	\end{equation}
	where $\psi=|\psi(\mathbf{r},t)|e^{i\theta(\mathbf{r},t)}$ is the order parameter field, 
	$\gamma$ is the relaxation rate,
	$(\phi,  \mathbf{A})$ is the electromagnetic scalar and vector potential,
	$\alpha\approx (T_{\text{c}}-T)/T_{\text{c}}$ is set by the temperature $T$ and the critical temperature $T_{\text{c}}$, $\xi_0$ is the coherence length,
	and we have set $2e=c=\hbar=1$.
	The electrical current can then be written in the `two-fluid' form as
	\begin{align}\label{eqn:current}
		\mathbf{j}=\mathbf{j}_{\mathrm{n}}+\mathbf{j}_{\mathrm{s}}=\hat{\sigma}_\mathrm{n} \mathbf{E} + \frac{n_{\mathrm{s}}}{m} (-\mathbf{A}+\nabla 
		\theta)
	\end{align}
	where $\hat{\sigma}_\mathrm{n}$ is the optical conductivity of normal carriers and $n_\mathrm{s} \propto |\psi|^2$ is the superfluid density.
	The dynamics implied by the relaxational TDGL equation is proved to be valid for superconductors rendered gapless by magnetic impurites~\cite{Gorkov1968}, or for superconducting fluctuations above the critical temperature $T_\mathrm{c}$~\cite{Cyrot1973, Larkin.2005_book, kamenev}. 
	For generic superconductors, TDGL serves as a simplifying approximation to the microscopic dynamics of coupled fermions and the order parameter field, which could not be captured by a local differential equation in time in terms of the order parameter itself~\cite{Cyrot1973, Larkin.2005_book,frank_incompatibility_2016}.
	The electrical current $\mathbf{j}$ in response to an electric field pulse $\mathbf{E}_0\delta(t)$ jumps to a nonzero value proportional to the carrier density  and then decays to zero due to momentum-relaxing scattering as $\sigma_\mathrm{n}(t,t')=\Theta(t-t^{\prime}) e^{-\gamma_i (t-t')} n_\mathrm{n}/m$ in a Drude metal, while it decays to a nonzero constant value proportional to the superfluid density $n_s$ in a superconductor, as shown by the gray and light blue curves in Fig.~\ref{fig:1}.
	
 In ultrafast experiments,  a pump laser pulse (typically within a picosecond) with a photon energy (infrared to visible range) much higher than the superconducting gap excites a lot of \red{quasiparticle}s, preparing them with a high effective temperature~\cite{demsar,suntt}. 
This suppresses the coefficient $\alpha$ in \equa{eq:tdgl} or even drives it negative, so that the order parameter evolves to a small value $\psi_0$. 
Subsequently, the electronic system cools down  via electron-phonon energy exchange,  restoring $\alpha$ to its original positive value. 
During the cooling stage, the dynamics of the order parameter can be simply captured by the mean-field TDGL equation linearized for small $\psi$, which yields an exponentially growing order parameter: $\psi(t) \sim \psi_0 e^{ \alpha \gamma t}$.
We investigate the linear optical conductivity of this nonequilibrium  state to a probe field at a frequency below the gap, typically within the terahertz (THz) range.
For simplicity, consider a static vector potential $\mathbf{A}$ as a probe field that already exists at time zero (a transverse vector potential that cannot be removed by gauge transformation), the initial current reads $\mathbf{j} =- n_\mathrm{s}\mathbf{A} /m=n_\mathrm{s}\mathbf{v}$ from the London equation.
During the  cooling stage,  the superconducting order recovers and the superfluid density grows while the flowing velocity does not change, which indicates an exponentially growing supercurrent $\mathbf{j} = n_\mathrm{s}(t)\mathbf{v}  \propto e^{2 \alpha \gamma t}$~\cite{kennes17,kenneser} shown by the dark blue curve in Fig.~\ref{fig:1}. 
Note that a spatial gradient of the phase ($\nabla \theta$) will not be induced by a transverse  vector potential at the linear response level, see \red{Appendix~\ref{appendix:TDGL_linear}}.
	
	This intriguing phenomenon may be called {\em supercurrent growth}, which is  \red{counterintuitive} since it appears to violate momentum conservation.
	Viewed in real space, as two free electrons combine to form a Cooper pair during the recovering dynamics, they intuitively obtain a total momentum of $2m\mathbf{v}$  before joining the stream of collectively flowing Cooper pairs, as shown by the inset of Fig.~\ref{fig:1}.
	In a Galilean invariant system where the bare electrons have the kinetic energy $\varepsilon_\mathbf{k}=\mathbf{k}^2/(2m)-\mu$, the total current $\mathbf{j}=\mathbf{P}/m$ is simply proportional to the total momentum
	which is conserved even in the presence of electron-electron interactions, 
	meaning that this phenomenon would be forbidden by momentum conservation.

	In the following, by resorting to microscopic theory, we show that supercurrent growth does indeed exist in real superconductors during ultrafast pump-probe experiments, and that the extra momentum is provided by momentum-relaxing scattering due to impurities and phonons. 
	The remainder of this paper is organized as follows. 
    In \red{Sec.}~\ref{section:Hamiltonian}, we introduce the Bardeen-Cooper-Schrieffer (BCS) Hamiltonian as the starting point. 
	In \red{Sec.}~\ref{section:Boltzmann}, we  describe the Boltzmann kinetic equation for the quasiparticle combined with the gap equation, and elucidate from it the supercurrent growth. 
	In \red{Secs.}~\ref{section:Meissener} and \ref{section:reflectivity}, we predict the experimental consequences of the supercurrent growth: the transient Meissner effect and the enhanced THz reflectivity. We conclude in \red{Sec.}~\ref{section:conclusion} with a brief summary and outlook.

	\section{Hamiltonian}
	\label{section:Hamiltonian}
    The microscopic Hamiltonian in the presence of a vector potential $\mathbf{A}$ can be expressed as $H=H_{\mathrm{BCS}}+H_{\mathrm{ee}}+H_{\mathrm{im}}+H_{\mathrm{ph}}$ where 
	\begin{equation}\label{eq:HBCS}
		H_{\mathrm{BCS}}=\sum_{\mathbf{k},s}\varepsilon_{\mathbf{k}-e\mathbf{A}}c^{\dagger}_{\mathbf{k},s}c_{\mathbf{k},s}
		+
		\sum_{\mathbf{k}}(\Delta c^{\dagger}_{\mathbf{k}\uparrow}c^{\dagger}_{-\mathbf{k}\downarrow}+ \text{h.c.})
	\end{equation}
	is the BCS mean field Hamiltonian.
	Here $\varepsilon_{\mathbf{k}}=\mathbf{k}^2/2m-\mu$ is the kinetic energy of free electrons represented by $c_{\mathbf{k},s}$ relative to the  chemical potential $\mu$ as a function of momentum $\mathbf{k}$, 
	$s\in\{\uparrow,\downarrow\}$ is the spin index,
	and $\Delta$ is the superconducting order parameter (gap). 
	In terms of the Bogoliubov \red{quasiparticle}s $\gamma_{\mathbf{k},s}$ with energy $E_{\mathbf{k}}=(\varepsilon_{\mathbf{k}}^2+|\Delta|^2)^{1/2}$, 
	one has 	$H_{\mathrm{BCS}}=\sum_{\mathbf{k},s}E_{\mathbf{k}}\gamma^{\dagger}_{\mathbf{k},s}\gamma_{\mathbf{k},s}$.
	The presence of the transverse vector potential locally dresses $E_{\mathbf{k}}$ to 
	the effective energy of Bogoliubov \red{quasiparticle}s:
	$\tilde{E}_{\mathbf{k}}= E_{\mathbf{k}}-(e/m)\mathbf{k}\cdot\mathbf{A}(\mathbf{r},t)$ to linear order in $\mathbf{A}$, which is now asymmetric in momentum space as shown in Fig.~\ref{fig:mm}(b). 
	Note that for a longitudinal $\mathbf{A}$, a nonzero $\nabla\theta$ can be induced at the linear order (see \append{appendix:TDGL_linear}) and what enters the dispersion would be the gauge invariant combination $e\mathbf{A}-\nabla \theta/2$~\cite{aronov}.
	$H_{\mathrm{im}/\mathrm{ph}/\mathrm{ee}}$ denotes the  electron-impurity, electron-phonon interactions, and the remaining electron-electron interactions not included by $H_{\mathrm{BCS}}$.  
	The order parameter is set by the gap equation
	\begin{align}\label{eq:gap_equation}
		\frac{1}{g}=\sum_{\mathbf{k}}\frac{1-2f_{\mathbf{k}}}{2 E_{\mathbf{k}}}
	\end{align}
	where $g$ is the s-wave attractive interaction strength and $f_{\mathbf{k}}$ is the distribution function  (occupation number) of Bogoliubov  \red{quasiparticle}s.
	The Hamiltonian $H$ together with the gap equation \eqref{eq:gap_equation} determines the coupled dynamics of the order parameter $\Delta(\mathbf{r},t)$ and \red{quasiparticle}s.
	
	The current operator is
	\begin{align}\label{eq:j}
		\hat{\mathbf{j}}=\hat{\mathbf{j}}_{\mathrm{p}}+\hat{\mathbf{j}}_{\mathrm{d}}
		,\quad
		\hat{\mathbf{j}}_{\mathrm{p}}=\frac{e}{m}\sum_{\mathbf{k},s}\mathbf{k}
		\gamma^{\dagger}_{\mathbf{k},s}\gamma_{\mathbf{k},s}
		,\quad
		\hat{\mathbf{j}}_{\mathrm{d}}=-\frac{\hat{n}e^2}{m}\mathbf{A}
	\end{align} 
	where $\hat{\mathbf{j}}_{\mathrm{p}}$ and $\hat{\mathbf{j}}_{\mathrm{d}}$ are known as the paramagnetic and the diamagnetic currents, respectively. 
	The expectation value of the diamagnetic current is $\langle\hat{\mathbf{j}}_{\mathrm{d}}\rangle=-ne^2\mathbf{A}/m$ with $n$ being the total electron density~\cite{coleman,altland}, which exists immediately after $\mathbf{A}$ is applied. 
	Note that the paramagnetic current contributed  by a Bogoliubov  \red{quasiparticle} is strictly $e\mathbf{k}/m$, regardless of its fractional charge and complicated group velocity.
	In the following, we will show how  microscopic scattering processes collaborate to affect the paramagnetic current and result in supercurrent growth.

\section{Boltzmann kinetic equation approach}
	\label{section:Boltzmann}
    The dynamics governed by the Hamiltonian  $H$ can be approximated by the Boltzmann kinetic  equation~\cite{aronov,gray}
	\begin{equation}\label{eq:fbe}
		\begin{split}
			&\quad\frac{\partial f_{\mathbf{k}}}{\partial t}+\frac{\partial\tilde{E}_{\mathbf{k}}}{\partial\mathbf{k}}\frac{\partial f_{\mathbf{k}}}{\partial\mathbf{r}}
			-\frac{\partial\tilde{E}_{\mathbf{k}}}{\partial\mathbf{r}}\frac{\partial f_{\mathbf{k}}}{\partial\mathbf{k}}
			=I_{\mathrm{im,ee,ph}}\left[f_{\mathbf{k}}\right]
		\end{split}
	\end{equation}
	for the distribution function  $f_{\mathbf{k}}(\mathbf{r},t)$
	of Bogoliubov  \red{quasiparticle}s.
	Here the spin indices are suppressed and the effects of  electron-impurity, electron-phonon, and electron-electron  scatterings are subsumed  in the collision integral $I_{\mathrm{im,ee,ph}}[f_{\mathbf{k}}(\mathbf{r},t)]$.
	The Boltzmann equation \eqref{eq:fbe} and the gap equation \eqref{eq:gap_equation} form a closed set of equations that depicts the incoherent nonequilibrium	dynamics of a superconductor. 
For superconductors, the Boltzmann equation   works for the slow and large scale dynamics of \red{quasiparticle}s with the characteristic time and length scales much larger than the inverse gap $1/\Delta$ (at the order of picoseconds) and  coherence length $\xi_0$ (tens to hundreds of nanometers)~\cite{aronov,gray}.	
It also neglects the quantum coherence of the pair excitation degrees of freedom, and thus does not contain 
		the information of collective modes of the order parameter~\cite{Sun.2020_collective}.
To quantify the incoherent evolution of \red{quasiparticle} distribution in the  time scale much longer than the inverse gap and in the larger-than-micrometer length scale, which is the case for the delay dynamics in typical pump-probe experiments, the Boltzmann equation is appropriate.

	
	
	We assume that the pump prepares the \red{quasiparticle}s in a non-thermal distribution with a high effective temperature $T_{\mathrm{e}}=T_{\text{H}}$ and a small gap determined by \equa{eq:gap_equation}, while  the phonons are at the environment temperature $T_{\mathrm{L}}$~\cite{demsar,Tinkham}.
	In the subsequent cooling process, the electron-phonon collision terms in \equa{eq:fbe} reduces the number and energy of \red{quasiparticle}s so that the gap recovers.
	
\begin{figure}[t]
		\centering    
		\includegraphics[width=\linewidth]{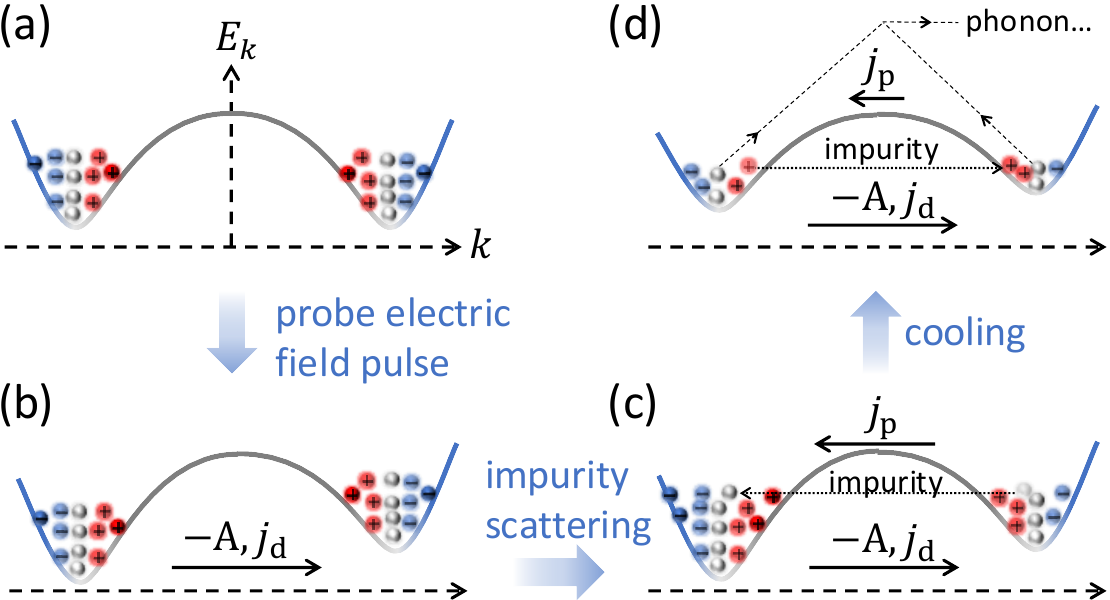}
		\caption{Microscopic mechanism for the current amplification. (a) Bogoliubov \red{quasiparticle} energy-momentum dispersion in the pump `heated' state with a high effective temperature $T_{\text{H}}$ and a  small gap. 
		There are  large amount of \red{quasiparticle}s represented by the spheres whose colors represent their charges. 
			(b) Right after the probe vector potential $\mathbf{A}(t)=-\mathbf{A}_0\Theta(t)$  is applied, there is an asymmetric \red{quasiparticle} dispersion and right-flowing  diamagnetic current $\mathbf{j}_{\mathrm{d}}$.
			(c)  Impurity scattering gives rise to the asymmetric  distribution $\delta f_{\mathbf{k}}^{\mathrm{a}}$ and the left-flowing  paramagnetic current $\mathbf{j}_{\mathrm{p}}$, partially canceling the diamagnetic current. 
			(d) Pair recombination due to phonon emission reduces the number  of \red{quasiparticle}s. 
			Impurities  scatter the \red{quasiparticle}s and relax the momentum imbalance  $\delta f_{\mathbf{k}}^{\mathrm{a}}$ and the paramagnetic current, increasing the total current.}
		\label{fig:mm}
	\end{figure}

	To derive the optical conductivity during the cooling stage, 
	we apply the probe  electric field pulse $\mathbf{E}(x,t)=-\partial_t \mathbf{A}(x,t)$ to a three or two dimensional system via the transverse vector potential $\mathbf{A}(x,t)=-\mathbf{A} e^{iq x} \Theta(t)$. Here $\mathbf A = A \hat{y}$ and the wavevector $q$ is taken along the $x$-direction without loss of generality. 
	To this end, one may separate the nonequilibrium distribution $f_{\mathbf{k}}(t)$ as
	\begin{align}\label{eq:fk}
		f_{\mathbf{k}}(t)= f^0_{\mathbf{k}}(t)+\delta f^{\mathrm{s}}_{\mathbf{k}}(t)+\delta f_{\mathbf{k}}^{\mathrm{a}}(t) e^{i\mathbf{q}\cdot \mathbf{r}}
	\end{align} 
	where $f_{\mathbf{k}}^0(t)=1/[e^{E_{\mathbf{k}}(t)/T_{\mathrm{L}}}+1]$  is the `equilibrium' distribution at time $t$ with $ E_{\mathbf{k}}(t)=(\epsilon_{\mathbf{k}}^2+|\Delta(t)|^2)^{1/2}$ being the  instantaneous \red{quasiparticle} energy.
	$\delta f^{\mathrm{s}}_{\mathbf{k}}(t)$ is a momentum-rotational-symmetric deviation  subject to pump-induced heating and electron-phonon cooling, also called the `energy mode'~\cite{Tinkham, demsar}. 
	$\delta f^{\mathrm{a}}_{\mathbf{k}}(t)\sim O(\mathbf{A})$ is a momentum-asymmetric deviation  subject to the probe field and momentum-relaxing scattering processes. Under the relaxation time approximation, the Boltzmann equation \eqref{eq:fbe} becomes
	\begin{subequations}\label{eq:berta}
		\begin{align}
			& \partial_t(f^0_{\mathbf{k}}+\delta f^{\mathrm{s}}_{\mathbf{k}})=-\gamma_{\mathrm{E}}\delta f^{\mathrm{s}}_{\mathbf{k}},
			\\
			&  
			\partial_t \delta f_{\mathbf{k}}^{\mathrm{a}}
			=-\left(\gamma_{\mathrm{i}}
			+i \mathbf{q}\cdot \mathbf{v}_{\mathbf{k}} \right)
			\left[
			\delta f^{\mathrm{a}}_{\mathbf{k}}
			-\frac{e}{m}\mathbf{k}\cdot\mathbf{A}
			\partial_{ E_{\mathbf{k}}}
			(f_{\mathbf{k}}^0+\delta f_{\mathbf{k}}^{\mathrm{s}})
			\right],
		\end{align}
	\end{subequations}
	where the first equation is at zeroth order in the probe field $\mathbf{A}$ and the second one is at the first order in it.
	For latter convenience, we included a nonzero wave vector $\mathbf{q}$ for the probe field, leading to the term $\mathbf{q}\cdot \mathbf{v}_{\mathbf{k}} $ with $ \mathbf{v}_{\mathbf{k}} =\partial_{\mathbf{k}} E_{\mathbf{k}}$ being the \red{quasiparticle} velocity. 
	$\gamma_{\mathrm{E}}$ is the energy relaxation rate of \red{quasiparticle}s due to electron-phonon scattering.
	It includes two kinds of processes: the intraband scattering of a \red{quasiparticle} by emitting/absorbing a phonon, and the `pair recombination/breaking' process that annihilates/creates two \red{quasiparticle}s  by emitting/absorbing an acoustic phonon as shown in \fig{fig:mm}(d).
	$\gamma_{\mathrm{i}}$ is the momentum relaxation rate
	\red{caused by} momentum-relaxing scattering processes including electron-phonon and electron-impurity scattering, see \fig{fig:mm}(c)(d). 
	In most materials, the electron-impurity scattering is the rapidest one at low temperature~\cite{aronov,demsar} and one has $\gamma_{\mathrm{E}}\ll\gamma_{\mathrm{i}}$. 
The electron-electron (e-e) scattering tends to thermalize the \red{quasiparticle} distribution but conserves the total momentum, current and \red{quasiparticle}  energy. 
		If $T_{\mathrm{e}}$ is lower than the critical temperature  $T_c$, the e-e scattering rate $\sim T_{\text{e}}^2/\varepsilon_{\text{F}}$ is in general very small.

	
	


\subsection{Equilibrium two-fluid model}
\label{section:Equilibrium_two_fluid}
    In an equilibrium superconductor, the \red{quasiparticle}s have a well defined temperature equal to that of the phonons: $T_{\mathrm{e}}=T_{\mathrm{L}}$. 
	Therefore, $\delta f^{\mathrm{s}}_{\mathbf{k}}$ in Eq.~\eqref{eq:berta} vanishes and the Boltzmann equation reduces to $\partial_t\delta f_{\mathbf{k}}^{\mathrm{a}}
	=-\gamma_{\mathrm{i}}\left[\delta f^{\mathrm{a}}_{\mathbf{k}}-\frac{e}{m}\mathbf{k}\cdot\mathbf{A}\partial_{ E_{\mathbf{k}}}f_{\mathbf{k}}^0\right]$ in the case of a uniform probe pulse, which yields the linear response  
	$\delta f_{\mathbf{k}}^{\mathrm{a}}=
	(1-e^{-\gamma_it})\frac{e}{m}\mathbf{k}\cdot\mathbf{A}\partial_{ E_{\mathbf{k}}}f_{\mathbf{k}}^0
	$.
	The  paramagnetic current is therefore obtained as $\mathbf{j}_{\mathrm{p}}=\frac{e}{m}\sum_{\mathbf{k}} 2\mathbf{k}
	\delta f_{\mathbf{k}}^{\mathrm{a}}
	=\mathbf{A}(1-e^{-\gamma_it}) n_{\mathrm{n}}e^2/m$ where 
	\begin{align} \label{eqn:nn_equi}
		n_{\mathrm{n}}=n \int d\epsilon  
		\left[-\partial_{E_{\mathbf{k}}} f_{0}\left( E_{\mathbf{k}}\right) \right]
		\equiv n_{\mathrm{n}}(\Delta, T_{\mathrm{L}})
		,\quad
		n_{\mathrm{s}}=n-n_{\mathrm{n}}
	\end{align}
	are defined as the `normal fluid density' and `superfluid density' 
	with $f_0( E_{\mathbf{k}})\equiv 1/[e^{E_{\mathbf{k}}/T_{\mathrm{L}}}+1]$ and $n$ being the total carrier density.
	Note that $n_{\mathrm{s}} \propto  |\Delta|^2/T_c^2$ at temperatures close to $T_{\mathrm{c}}$, and increases monotonically to $n$ as the temperature decreases to zero for a clean superconductor~\cite{Tinkham,altland}.
	Adding the diamagnetic current $\mathbf{j}_{\mathrm{d}}=-\mathbf{A}ne^2/m$, one arrives at the two-fluid model in \equa{eqn:current} with the Drude component $\sigma_\mathrm{n}(t,t')=\Theta(t-t^{\prime}) e^{-\gamma_i (t-t')} n_\mathrm{n}/m$.
	In the long time limit,  the only remaining current will be $\mathbf{j}=-n_{\mathrm{s}}e^2\mathbf{A}/m$, as shown by the light blue curve in Fig.~\ref{fig:1}.


\subsection{Nonequilibrium two-fluid model: supercurrent growth}
\label{section:Supercurrent growth}
    We now discuss the case of a nonequilibrium superconductor, which is the central result of this paper.
	Before computing the linear current response to the probe field, we first establish the picture of the cooling process of the nonequilibrium distribution function 
	$
	f^s_{\mathbf{k}}(t)= f^0_{\mathbf{k}}(t)+\delta f^{\mathrm{s}}_{\mathbf{k}}(t)
	$.
	Its time evolution could be obtained from Eq.~(\ref{eq:berta}a) with the time-dependent gap $\Delta(t)$ updated self consistently following \equa{eq:gap_equation}.	
	During the cooling dynamics, the  \red{quasiparticle} occupation 
	$
	f^s_{\mathbf{k}}(t)
	$
	decays with the rate $\gamma_{\mathrm{E}}$ due to electron-phonon scattering (mostly through the `pair recombination' process) toward $f^0_{\mathbf{k}}(t)$, the Fermi distribution at the lattice temperature but with the instantaneous dispersion $E_{\mathbf{k}}(t)$. 
	Meanwhile, the order parameter $\Delta(t)$ increases according to the gap equation, generating an additional  $\delta f^{\mathrm{s}}_{\mathbf{k}}$ following Eq.~(\ref{eq:berta}a). 
	See \append{appendix:solution_Boltzmann} for typical numerical solutions of $\Delta(t)$.
	Assuming the time dependence of $\Delta(t)$ is known, the distribution function can be expressed as
	\begin{align}\label{eq:fsk}
		\delta f^{\mathrm{s}}_{\mathbf{k}}(t)
		=& -
		\int_{0}^t dt' e^{\gamma_{\text{E}}(t'-t)}  
		\partial_{t'}  f^0_{\mathbf{k}}(t')
		+ e^{-\gamma_{\text{E}}t}  \delta f^{\mathrm{s}}_{\mathbf{k}}(0)
		\notag \\     
		 \xrightarrow{\gamma_{\text{E}} \gg \gamma_{\Delta},\,\,\, t\gg 1/\gamma_{\text{E}}} 
		 &
		-\frac{1}{\gamma_{\text{E}}} 	\partial_{t}  f^0_{\mathbf{k}}(t)
		=-\frac{1}{\gamma_{\text{E}}} 	\partial_{t}f_{0}\left( E_{\mathbf{k}}(t)\right)
		\,.
	\end{align}
	
	On top of the symmetric nonequilibrium distribution $f^{\mathrm{s}}_{\mathbf{k}}(t)$, one may apply the spatially uniform probe field $\mathbf{A}(t)=-\mathbf{A}\Theta(t)$ and compute its linear response $\delta f_{\mathbf{k}}^{\mathrm{a}}$ from Eq.~(\ref{eq:berta}b). 
	From it one obtains the paramagnetic current 
	\begin{align}\label{eq:tdpc}
		\mathbf{j}_{\mathrm{p}}(t)
		=& \mathbf{A}(t)
		\int_{0}^t dt' \gamma_i e^{\gamma_i(t'-t)}          n_{\mathrm{n}}(t')
		\notag \\     
		\xrightarrow{\gamma_i \gg \gamma_{\Delta}} 
		&
		\mathbf{A}(t) (1-e^{-\gamma_{\mathrm{i}}t})\frac{n_{\mathrm{n}}(t)e^2}{m}
	\end{align}
and the total current in the form of a two-fluid model:
	\begin{equation}\label{eq:tdtc}
		\mathbf{j}(t)=-[n_{\mathrm{s}}(t)+n_{\mathrm{n}}(t)e^{-\gamma_{\mathrm{i}}t}]\frac{e^2}{m}\mathbf{A}(t)
		\,.
	\end{equation}
	Here we define  the nonequilibrium normal fluid and superfluid densities as the natural generalizations of the equilibrium case in \equa{eqn:nn_equi}:
	\begin{equation}
		\begin{split}\label{eq:nn}
			n_{\mathrm{n}}(t)
			&=
			n \int d\epsilon  
			\left[-\partial_{E_{\mathbf{k}}} f^{\mathrm{s}}_{\mathbf{k}}(t) \right]  
			,\quad
			n_{\mathrm{s}}=n-n_{\mathrm{n}}
			,
		\end{split}
	\end{equation}
	with the equilibrium distribution function replaced by the transient one $f^{\mathrm{s}}_{\mathbf{k}}(t)$.
	In the case of $\gamma_{\text{E}} \gg \gamma_{\Delta}$, one may compute the leading order dynamical correction to the normal fluid density from  \equa{eq:fsk} and \equa{eq:nn} as
	\begin{equation}
		\begin{split}\label{eq:nn2}
			n_{\mathrm{n}}(t)
			&=
			n_{\mathrm{n}}\left[\Delta(t), T_{\mathrm{L}}\right]
			+n \frac{\partial_{t}\Delta}{\gamma_{\text{E}}}  \int d\epsilon  
			\left[\partial_{E_{\mathbf{k}}} \frac{\Delta}{E_{\mathbf{k}}}	\partial_{E_{\mathbf{k}}} 
			f_{0}\left( E_{\mathbf{k}}\right)\right]  
			\,.
		\end{split}
	\end{equation}
	It reduces to the equilibrium one in the static limit. 
	Therefore, as the gap recovers in the cooling process, the superfluid density $n_{\mathrm{s}}(t)$ increases since the normal carrier density $n_{\mathrm{n}}(t)$ decreases according to the first term in \equa{eq:nn2}. 
Note that the second term in \equa{eq:nn2} is a positive dynamical correction to the normal fluid density used in TDGL.

	%

	The physical picture of this response is shown in \fig{fig:mm}(b)(c).
	Because of the vector potential, the \red{quasiparticle} energy is corrected to the  asymmetric form
	$\tilde{E}_{\mathbf{k}}= E_{\mathbf{k}}-(e/m)\mathbf{k}\cdot\mathbf{A}$   in momentum space  shown in Fig.~\ref{fig:mm}(b). 
	The elastic scattering from impurities thus scatters quasiparticles from right to left following Fig.~\ref{fig:mm}(b)(c), evolving the quasiparticle distribution from $	f^s( E_{\mathbf{k}})$ to $f^s(\tilde{E}_{\mathbf{k}})$, whose difference gives the last term in Eq.~(\ref{eq:berta}b). 
	Therefore, an asymmetric component  $\delta f_{\mathbf{k}}^{\mathrm{a}}$ emerges with the rate $\gamma_i$ as shown in Fig.~\ref{fig:mm}(c). 
	This contributes a nonzero paramagnetic current flowing to the left, partially canceling the diamagnetic current $\mathbf{j}_{\mathrm{d}}=\mathbf{A} ne^2/m$ flowing to the right.
	Within the time scale $1/\gamma_i$, the paramagnetic current  grows to $-\mathbf{A} n_{\mathrm{n}}(t)e^2/m$ so that the total current  decays to $\mathbf{j}=\mathbf{A} n_{\mathrm{s}}(t)e^2/m$ which flows to the right, as shown in Fig.~\ref{fig:mm}(c). 
	
	Afterwards, the total current grows with the superfluid density $n_{\mathrm{s}}(t)$ following \equa{eq:tdtc}, giving rise to \emph{supercurrent growth}.
	The microscopic mechanism is depicted in Fig.~\ref{fig:mm}(c)(d). 
	As the energy and number of \red{quasiparticle}s  decrease due to electron-phonon scattering, e.g., two \red{quasiparticle}s annihilate by the emission of a phonon as in Fig.~\ref{fig:mm}(d),  the gap recovers according to the gap equation. 
	Imagine  the gap becomes so large in the end  that very few \red{quasiparticle}s are left, the paramagnetic current flowing to the left must have decayed (see inset of \fig{fig:1}), so that the total current  increases.
	However, a momentum conserved annihilation of two \red{quasiparticle}s with momenta $\mathbf{k}$ and $-\mathbf{k}$ cannot change the paramagnetic current 
	$\mathbf{j}_{\mathrm{p}}=\frac{e}{m}\sum_{\mathbf{k}} 2\mathbf{k}
	f_{\mathbf{k}}
	$. 
	In fact, the momentum of the emitted phonon is so small due to the small gap in realistic superconductors that the momenta of the annihilated two \red{quasiparticle}s are roughly opposite to each other.
	What actually increases the total current is the momentum relaxing scattering processes, such as the elastic scattering of \red{quasiparticle}s from left to right by impurities  shown in Fig.~\ref{fig:mm}(d).  In this way, the paramagnetic current decreases so that the total current flowing to the right increases.
	Therefore, we emphasize that   increasing of the gap alone {\em does not} lead to current amplification. 
	Indeed, \equa{eq:tdpc} shows that in the  limit of small $\gamma_i$, the paramagnetic current is zero and won't follow the instantaneous normal fluid density.

	The nonequilibrium optical conductivity 
	defined by $\mathbf{j}(t)=\int dt^{\prime}\sigma_{\text{s}}(t,t^{\prime})\mathbf{E}(t^{\prime})$ thus reads
	\begin{equation}\label{eq:mfoc}
		\sigma_{\text{s}}(t,t^{\prime})= \frac{e^2}{m}  n_{\mathrm{s}}(t) \theta(t-t^\prime)
	\end{equation}	
	for $t-t^{\prime} \gg \tau_i= 1/\gamma_i$.
The Boltzmann equation \eqref{eq:fbe} and the gap equation \eqref{eq:gap_equation} predict a negative exponential recovery 
of the superfluid density 
$
n_{\mathrm{s}}(t) \approx n_{s0} +[n_s(0)-n_{s0}]e^{- \gamma_\Delta t}
$ 
towards its equilibrium value $n_{s0}$, see Figs.~\ref{fig:1} and \ref{fig:gap_relaxation}. 
This constitutes a qualitative modification of the early-time exponential  growth 
$n_\mathrm{s}(t) \propto e^{2 \alpha \gamma t}$ 
predicted by TDGL theory.
At temperatures near $T_{\mathrm{c}}$, 
the relaxation rate is the Schmid-Sch\"{o}n rate $\gamma_{\Delta}\approx 0.27 \gamma_{\mathrm{E}}\Delta/T_{\mathrm{c}}$~\cite{schmid_approach_1968, gray,Tinkham}. 
The recovery time scale $1/\gamma_{\Delta}$ ranges from $10 \unit{ps}$ to $10 \unit{ns}$ in conventional superconductors, while
the phonon bottleneck effect~\cite{Rothwarf.1967,demsar, Zhao.2024_phonon_bottleneck} not included in \equa{eq:berta}  may further slow down the cooling dynamics.
The recovery time was found to be as short as a few picoseconds in cuprates~\cite{Averitt.2001}.
One could measure $\sigma_{\text{s}}(t,t^{\prime})$ in this time window by the reflectivity of a probe pulse.

	To summarize the analysis above, the \red{counterintuitive} phenomenon of
	supercurrent growth  indicated by the TDGL equation is confirmed from the Boltzmann equation. We  conclude that {\em impurities along with energy relaxation through phonons} underlie the  microscopic mechanism for current amplification, contrary to the common belief that impurities always attenuate currents.

\section{Transient Meissner effect}
\label{section:Meissener}
\begin{figure}
		\centering    \includegraphics[width= \linewidth]{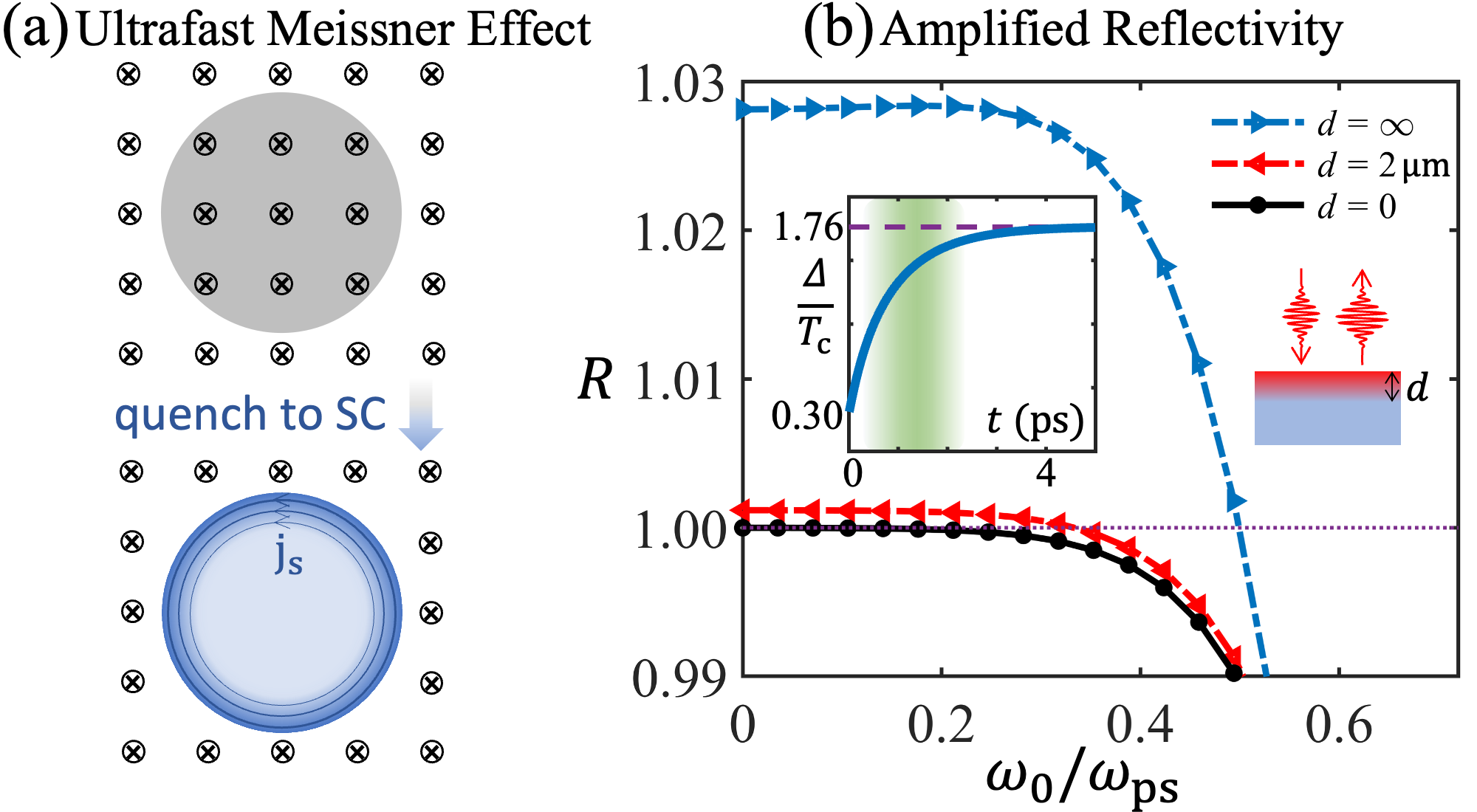}
		\caption{
			(a) Illustration of the ultrafast Meissner effect during the quench to a superconducting state. The magnetic fluxes (black crosses) are 	expelled out of the sample as the supercurrent (blue lines) grows in time.
			(b) Reflectivity of a normally incident Gaussian light pulse $E(t,z)=E_0 e^{-(t-t_0+z/c)^2/(2\delta^2)-i\omega_0 (t+z/c)}$ on a nonequilibrium dirty superconductor (right inset)  for three different penetration depths $d$.
			The time dependence of the gap is shown by the left inset, which is obtained from solving Eqs.~(\ref{eq:berta}a) and \eqref{eq:gap_equation} with $\gamma_{\text{E}}=1 \unit{THz}$, $T_{\mathrm{L}}=0.20\, T_{\text{c}}$ and $T_{\mathrm{H}}=0.99 \, T_{\text{c}}$. 
			As a result, the superfluid plasma frequency grows from $0.41 \,\omega_{\mathrm{ps}}$ at $t=0 \unit{ps}$ to approach the equilibrium one  $\omega_{\mathrm{ps}}=\sqrt{4\pi n_{s0} e^2 /m}=1.7 \unit{THz}$ after  $t=4 \unit{ps}$.
			The red stripe in the left inset marks the central time and  width $\delta= 0.3 \unit{ps}$ of the light pulse on the surface of the sample.
			The  choice of parameters is motivated by those of YBCO~\cite{Averitt.2001,Hoegen.2022}.
		}  \label{fig:rf}
\end{figure}

An immediate physical consequence of the supercurrent growth is the transient Meissner effect. 
During the quench of a metal into a superconductor in the presence of a static magnetic field, if the Meissner effect establishes,
the supercurrent $\mathbf{j}(t)=-n_{\mathrm{s}}(t)e^2\mathbf{A}/m$ must have grown in time with the superfluid to expel the magnetic field toward the outside of the sample~\cite{Hirsch.2016,Hirsch.2017}, see \fig{fig:rf}(a). 
To prove this current response, one may simply solve the Boltzmann equation in \equa{eq:berta} in a transverse vector potential  $\mathbf{A}(\mathbf{r})=\mathbf{A}e^{iqx}$  that represents the magnetic field $\mathbf{B}=\nabla \times \mathbf{A}(\mathbf{r})$. 
Note that because of the nonzero wave vector of the field and the response current, the global momentum conservation no longer forbids the local current growth, and the supercurrent may grow due to the exchange of momenta between upward and downward current streams from the $qv_{\mathbf{k}x}$ term in \equa{eq:berta},  even without the help of impurities or phonons.
Nevertheless, in macroscopic samples of size $L\sim \unit{cm}$ under a nearly uniform magnetic field, the characteristic wave vector of the vector potential is $q \sim 1/L$, leading to the typical $qv_{\mathbf{k}x}\sim 0.1 \unit{GHz}$ much smaller than the typical momentum relaxation rate $\gamma_i \sim\unit{THz}$.
Thus it is the $\gamma_i$ term that dominates in \equa{eq:berta} which yields Eqs.~\eqref{eq:tdtc}-\eqref{eq:nn}  for the current amplification.
Therefore, 
if the ultrafast Meissner effect happens in a macroscopic sample, it relies microscopically on the electron-impurity and electron-phonon scattering in \fig{fig:mm}. 
This picture could be tested by measuring the time scales of suppression and reestablishment of the Meissner effect in an ultrafast experiment in conventional superconductors.

Signatures of transient expulsion of magnetic field were observed by a recent experiment~\cite{fava}, where 
the scenario of light-induced superconductivity \emph{above} its transition temperature was suspected.
Regardless of the exact mechanism of light-induced superconductivity, 
as the supercurrent must grow in time to expel the magnetic field, 
this growth should arise microscopically from impurities/phonons as in Fig.~2 or from other current-nonconserving scattering processes.



\section{Enhanced THz reflectivity}
\label{section:reflectivity}
Because supercurrent growth of a nonequilibrium superconductor implies that the transient state acts as a gain medium, 
it may lead to  enhanced reflectivity of light ~\cite{Rajasekaran2016Parametric, mitrano16,cantaluppi_pressure_2018,buzzi,rowe23,Averitt.2001, Hoegen.2022}.
This is confirmed by the numerical results of the reflectivity shown in Fig.~\ref{fig:rf}.
	To model the finite penetration depth $d$ of the pump, we use the spatial  profile 
$\sigma\left(z; t,t^{\prime}\right)=e^{z/d}\sigma_{\mathrm{s}}(t,t^{\prime})+(1-e^{z/d})  \Theta(t-t^{\prime})  n_{s0}e^2/m$ for the optical conductivity. 
It interpolates from the nonequilibrium surface with $\sigma_{\mathrm{s}}$ from \equa{eq:mfoc} at $z=0$ to the equilibrium superconducting bulk at $z \ll -d$, see right inset of Fig.~\ref{fig:rf}(b).
Because the reflectivity is computed for frequencies below the gap of a dirty superconductor with $\gamma_{\text{i}} \gg \Delta$, the normal fluid contribution is neglected.
One finds that the reflectivity is enhanced (red and blue lines) compared to  the equilibrium reflectivity (black line).
Notably, 
the reflectivity exceeds unity for a range of frequencies which is forbidden in equilibrium systems.

The larger-than-unity reflectivity could be understood intuitively in the time domain.
When the probe pulse arrives at the interface (see left inset of \fig{fig:rf}(b)), the plasma frequency $\omega_{\text{ps}}(t) \sim \sqrt{n_s(t)}$ of the superconductor is still below the central frequency $\omega_0$ of the pulse, so that part of the pulse could penetrate into the surface layer of the superconductor. 
The penetrated pulse gets adiabatically amplified~\cite{Sun.2016} as it propagates
in the pumped layer which is a gain medium.
However, the amplified  pulse cannot keep penetrating into the superconductor's deep interior for two reasons. 
Firstly, the plasma frequency $\omega_{\text{ps}}(t)$ will increase to above the pulse frequency as time goes on.
Secondly, the plasma frequency $\omega_{\text{ps}}$ deep in the sample is in equilibrium and is higher than the pulse frequency.
Therefore, the EM energy generated by the gain cannot penetrate into the superconductor's interior and must be emitted out of the sample, leading to a reflectivity $R > 1$. 
Appendices~\ref{appendix:Enhanced_reflectivity} and \ref{appendix:perturbation} contain analytical understanding of this effect.
\section{Conclusion}
\label{section:conclusion}
In summary, we have elucidated  that the mechanism of supercurrent growth relies crucially on the  electron-impurity and electron-phonon scattering.
This  changes the common intuition that impurities may only  attenuate the current. 
A nonequilibrium two-fluid optical conductivity is derived from the Boltzmann kinetic equation in \red{Sec.}~\ref{section:Supercurrent growth}.
We further predict its interesting consequences 
relevant to existing experiments: the ultrafast Meissner effect~\cite{fava}, and an amplified optical reflectivity exceeding unity~\cite{buzzi,Hoegen.2022}.
We emphasize that in the time-domain response to a delta-function electric field,  the supercurrent growth terminates as the superfluid density approaches the equilibrium value $n_{s0}$, and thus cannot exceed  $\mathbf{j}_0=-n_{s0} \mathbf{A}e^2/m$.
For a pulse whose electric field integrates to zero~\cite{kennes17}, the supercurrent even vanishes after the pulse.
However, this limitation does not hinder its intriguing transient  consequences.
For example, it leads to amplification of EM waves (\fig{fig:rf}(b)) and other collective modes~\cite{Sun.2016} such as the superfluid plasmons~\cite{Sun.2020_collective,  Gabriele_non-linear_2021, Sellati_Josephson_plasmon.2023, Sellati_Josephson_plasmon.2025, Zhang_2D_spectroscopy.2023,  Salvador_2D_spectroscopy.2024, taherian_probing_2025} and the Carlson-Goldman mode~\cite{Carlson.1975, Sun.2020_collective}, which is not possible in equilibrium.
	These effects, including the related phenomena in fluctuation dominated nonequilibrium superconductors containing phase slips~\cite{kenneser, Lemonik.2018, suntt, Eckstein.2021, Wang.2021},  are left for future research.

	
	%

	\begin{acknowledgements}
		Q.Y., Y.C. and Z.S. are supported by  the National Natural Science Foundation of China (Grants No. 12421004 and No. 12374291),  the National Key	Research and Development Program of China (2022YFA1204700),  Beijing Natural Science Foundation (Z240005),  and the startup grant from Tsinghua University. 
		D.M.K acknowledges funding by the Deutsche Forschungsgemeinschaft (DFG, German
			Research Foundation) - 508440990 - 531215165 (Research Unit ``OPTIMAL'').
		We thank Yi Zuo and Can Huang for helpful discussions. 
	\end{acknowledgements}

$^*$ These authors contributed equally to this work.

$^\dagger$ Corresponding author:  zysun@tsinghua.edu.cn

    \section*{Data Availability}
    The data and codes that support the  gap evolution and optical reflectivity  presented in this study are openly available~\cite{ProjectData}.


\appendix

\section{Derivation of the linearized Boltzmann equation}
In this section, we derive the linearized Boltzmann equation in \equa{eq:berta} by plugging the trial distribution function in \equa{eq:fk} into the original Boltzmann equation in \equa{eq:fbe}.
	Note that the probe field enters \equa{eq:fbe} via the transverse vector potential $\mathbf{A}(\mathbf{r},t)$ in the space-momentum dependent \red{quasiparticle} energy $\tilde{E}_{\mathbf{k}}= E_{\mathbf{k}}-(e/m)\mathbf{k}\cdot\mathbf{A}(\mathbf{r},t)$.

	At zeroth order in $\mathbf{A}$, the Boltzmann equation describes the relaxation of the nonequilibrium background that involves only the momentum symmetric distributions $f^0_{\mathbf{k}}(t)$ and $\delta f^{\mathrm{s}}_{\mathbf{k}}(t)$, which are space-independent.
	Therefore, the $\partial_{\mathbf{r}} f_{\mathbf{k}}$ term  vanishes in \equa{eq:fbe}.
	Moreover, the $\partial_{\mathbf{r}} \tilde{E}_{\mathbf{k}}$ term is at linear order in $\mathbf{A}$ and must drop out too.
	What remains is the first term in \equa{eq:fbe} and the collision term on its right hand side: $\partial_t(f^0_{\mathbf{k}}+\delta f^{\mathrm{s}}_{\mathbf{k}})=I_{\mathrm{im,ee,ph}}[f]$.
	In the collision term, scattering of \red{quasiparticle}s from impurities does not relax momentum symmetric distribution, and thus has no effect at this order.
	Inter particle scattering and scattering from phonons act to thermalize the \red{quasiparticle} distribution to a Fermi-Dirac one at the temperature $T_{\mathrm{L}}$ of the lattice.
	It  thus vanishes for
	$f_{\mathbf{k}}=f_{\mathbf{k}}^0(t)=1/[e^{E_{\mathbf{k}}(t)/T_{\mathrm{L}}}+1]$ and would have an effect only if there is a nonzero `energy mode' $\delta f^{\mathrm{s}}_{\mathbf{k}}$.
	Therefore, at the level of relaxation time approximation with the energy relaxation rate $\gamma_{\mathrm{E}}$, one has $I_{\mathrm{im,ee,ph}}[f]=-\gamma_{\mathrm{E}}\delta f^{\mathrm{s}}_{\mathbf{k}}$ which yields  Eq.~(\ref{eq:berta}a).

	We now proceed to derive Eq.~(\ref{eq:berta}b).
	At linear order in $\mathbf{A}$, the probe field induces a small `momentum mode' $\delta f_{\mathbf{k}}^{\mathrm{a}}(t) e^{i\mathbf{q}\cdot \mathbf{r}}$ that is asymmetric in momentum $\mathbf{k}$. 
	Therefore, momentum-relaxing scattering from impurities and phonons would relax this component with the rate $\gamma_i \delta f_{\mathbf{k}}^{\mathrm{a}}$.
	Furthermore, momentum-relaxing scattering conserves the \red{quasiparticle} energy $\tilde{E}_{\mathbf{k}}= E_{\mathbf{k}}-(e/m)\mathbf{k}\cdot\mathbf{A}(\mathbf{r},t)$ that is tilted in momentum space, so that it tends to generate a momentum-asymmetric distribution out of the symmetric distribution $f^0_{\mathbf{k}}+\delta f^{\mathrm{s}}_{\mathbf{k}}$, see \fig{fig:mm}(b)(c).
	At linear order in $\mathbf{A}$, this generation rate is thus $\gamma_i (e/m)\mathbf{k}\cdot\mathbf{A} \partial_{E_{\mathbf{k}}}
	(f^0_{\mathbf{k}}+\delta f^{\mathrm{s}}_{\mathbf{k}})$.
	The energy relaxing scattering has no effect in the momentum-assymmetric deviation of distribution because the latter does not alter the total energy at linear order in $\mathbf{A}$. 
	Adding these two terms, the collision term reads
	\begin{align}
		I_{\mathrm{im,ee,ph}}[f]=-\gamma_{i}
		\left[\delta f^{\mathrm{a}}_{\mathbf{k}} -
		\frac{e}{m}\mathbf{k}\cdot\mathbf{A} \partial_{E_{\mathbf{k}}}(f^0_{\mathbf{k}}+\delta f^{\mathrm{s}}_{\mathbf{k}})
		\right]
	\end{align} 
	where we have removed the overall spatial dependence $e^{i\mathbf{q}\cdot \mathbf{r}}$ for notational simplicity.
	At $O(\mathbf{A})$, the first term in  \equa{eq:fbe} is simply $\partial_t \delta f_{\mathbf{k}}^{\mathrm{a}}$.
	The second term reads $\frac{\partial\tilde{E}_{\mathbf{k}}}{\partial\mathbf{k}}\frac{\partial f_{\mathbf{k}}}{\partial\mathbf{r}}=\mathbf{v}_{\mathbf{k}} \cdot  i \mathbf{q} \delta f_{\mathbf{k}}^{\mathrm{a}}$.
	The third term is
	\begin{align}
		\frac{\partial\tilde{E}_{\mathbf{k}}}{\partial\mathbf{r}}\frac{\partial f_{\mathbf{k}}}{\partial\mathbf{k}} 
		&=-\frac{e}{m}(\mathbf{k}\cdot\mathbf{A}) 
		\left[
		i\mathbf{q}\cdot \partial_{\mathbf{k}} (f^0_{\mathbf{k}}+\delta f^{\mathrm{s}}_{\mathbf{k}})
		\right]
		\notag\\
		&=
		-\frac{e}{m}(\mathbf{k}\cdot\mathbf{A}) 
		\left[
		i(\mathbf{q}\cdot 
		\mathbf{v}_{\mathbf{k}})
		\partial_{ E_{\mathbf{k}}}
		(f^0_{\mathbf{k}}+\delta f^{\mathrm{s}}_{\mathbf{k}})
		\right]
	\end{align}
	where we have made use of 
	$\partial_{\mathbf{k}} = (\partial_{\mathbf{k}} E_{\mathbf{k}})
	\partial_{ E_{\mathbf{k}}} = \mathbf{v}_{\mathbf{k}}
	\partial_{ E_{\mathbf{k}}}$.
	Adding these three terms and the collision term, one arrives at Eq.~(\ref{eq:berta}b).

At $q=0$, the solution to Eq.~(\ref{eq:berta}b) is 
\begin{align}
\delta f_{\mathbf{k}}^{\mathrm{a}}(t)
&=
\int_0^t dt'
\gamma_i 
\frac{e}{m}\mathbf{k}\cdot\mathbf{A}(t')
\partial_{ E_{\mathbf{k}}}
f^s_{\mathbf{k}}(t')
\,.
\end{align}
The integral $\mathbf{j}_{\mathrm{p}}=\frac{e}{m}\sum_{\mathbf{k}} 2\mathbf{k}
\delta f_{\mathbf{k}}^{\mathrm{a}}
$ yields the paramagnetic current in \equa{eq:tdpc}.

\section{Dirty superconductors}
So far, our analysis using the kinetic equation is limited to superconductors whose momentum relaxing scattering rate $\gamma_{\mathrm{i}}$ is much smaller than the critical temperature $T_{\text{c}}$, which are called clean superconductors.
For dirty superconductors where $\gamma_{\mathrm{i}} \gg T_{\text{c}}$, 
the superfluid density is not equal to the total carrier density even at zero temperature, but smaller than it by a ratio $\pi  |\Delta|/\gamma_{\mathrm{i}}$~\cite{mattis1958theory}.
In this case, the same supercurrent growth should exist in nonequilibrium as long as the gap increases in time.
	
This  supercurrent growth  is verified by the numerical results in Ref.~\cite{kennes17} (see Fig.~8 there) for a dirty superconductor, which  considers a temporal profile of the gap $\Delta(t)$ that rises from zero and then either stays at a non-zero value or falls back to zero. 
The nonequilibrium optical conductivity was computed using Green functions with the Mattis-Bardeen approximation~\cite{mattis1958theory}.
Furthermore, the nonequilibrium optical conductivity in the frequency domain (Fig.~15 of Ref.~\cite{kennes17}) is qualitatively similar to the Fourier transform of \equa{eq:tdtc} for the clean case. 
Crucially, the real part of the optical conductivity of the dirty  nonequilibrium superconductor can turn negative, indicating an instability in the time domain. 
Again, the supercurrent growth occurs there due to impurity scattering, 
which is contained implicitly in the  Mattis-Bardeen approximation that completely relaxes the constraint of momentum conservation.
For frequency scales much smaller than the gap, it is enough to keep the superfluid response which yields \equa{eq:mfoc} with the superfluid density replaced by that for a  Dirty superconductor: $n_{\mathrm{s}}(t)=n \pi \Delta(t)/\gamma_i $.

\begin{figure}
	\centering    \includegraphics[width= 0.9 \linewidth]{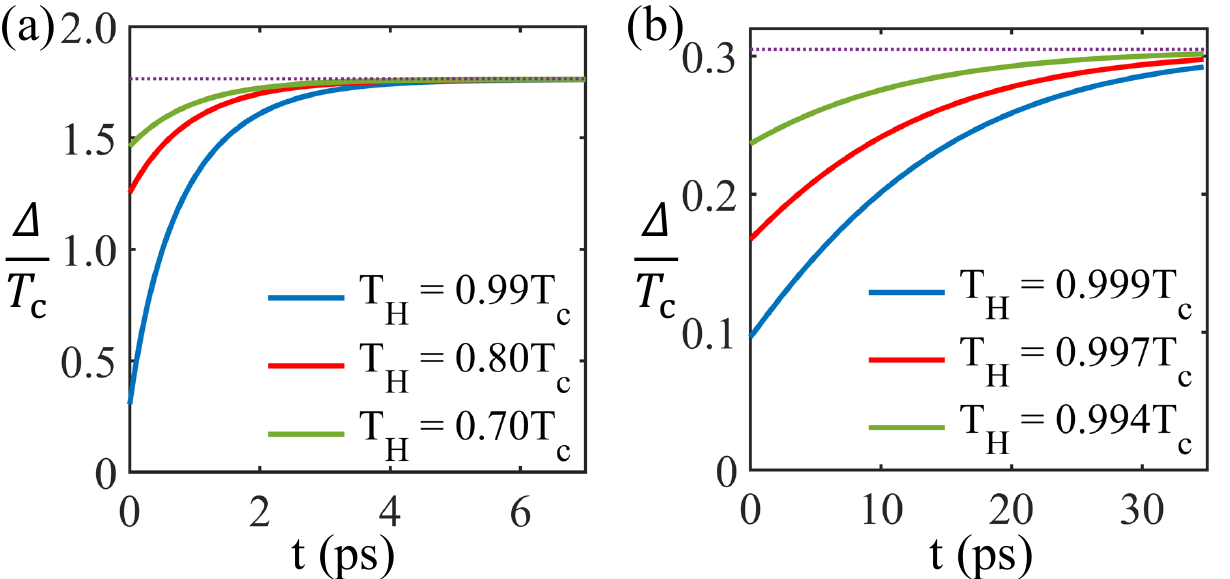}
	\caption{
		Simulated relaxation dynamics of the energy gap  from  Eqs.~(\ref{eq:berta}a) and \eqref{eq:gap_equation} with $\gamma_{\text{E}}=1 \unit{THz}$ and different initial \red{quasiparticle} temperatures $T_{\mathrm{H}}$.
		The lattice temperature is fixed at (a) $T_{\mathrm{L}}=0.20 \, T_{\mathrm{c}}$ and (b) $T_{\mathrm{L}}=0.990 \, T_{\mathrm{c}}$.
	}  \label{fig:gap_relaxation}
\end{figure}

\section{Numerical solution of the Boltzmann equation}
\label{appendix:solution_Boltzmann}
To verify the relaxational dynamics for the gap during electron-phonon cooling, 
we numerically simulated the time evolution of the order parameter $\Delta$ by solving Eq.~(\ref{eq:berta}a) coupled with  Eq.~\eqref{eq:gap_equation}. 
At $t = 0$, we assume the quasiparticle distribution corresponds to a thermal  state at an elevated temperature $T_\mathrm {H}$,  while the lattice has a fixed temperature $T_{\mathrm{L}}$ that enters $f_{\mathbf{k}}^0(t)=1/[e^{E_{\mathbf{k}}(t)/T_{\mathrm{L}}}+1]$ in Eq.~(\ref{eq:berta}a).
Typical numerical results are shown in \fig{fig:gap_relaxation}, which demonstrates that the gap grows from its initial value to the equilibrium value $\Delta_{\text{L}}$ set by $T_{\mathrm{L}}$, roughly in an exponential way.
We define the gap relaxation time $\tau_{\Delta}(t)$ for the gap dynamics by
$
d \Delta(t)/dt = -[\Delta(t) - \Delta_{\text{L}}]/\tau_{\Delta}(t)
	\label{eq:gaprelax}
$
and extract  $\tau_{\Delta}(t)$ from the numerical result of $\Delta(t)$.
For $T_\mathrm {H},\, T_\mathrm {L} \ll T_\mathrm{c}$,  the gap relaxation rate approaches the \red{quasiparticle} energy relaxation rate $\gamma_\mathrm E$.
Near the critical temperature so that
$T_\mathrm {H},\, T_\mathrm {L} \approx T_\mathrm{c}$, 
 the gap relaxation exhibits a much longer timescale, which agrees with the Schmid-Sch\"{o}n relaxation time $\tau_{\Delta}=3.7 \mathrm {T_c}/(\Delta \gamma_{\mathrm{E}} )$~\cite{schmid_approach_1968, gray,Tinkham}.

\section{Enhanced reflectivity}	
\label{appendix:Enhanced_reflectivity}
\begin{figure}
	\centering    \includegraphics[width=0.5 \linewidth]{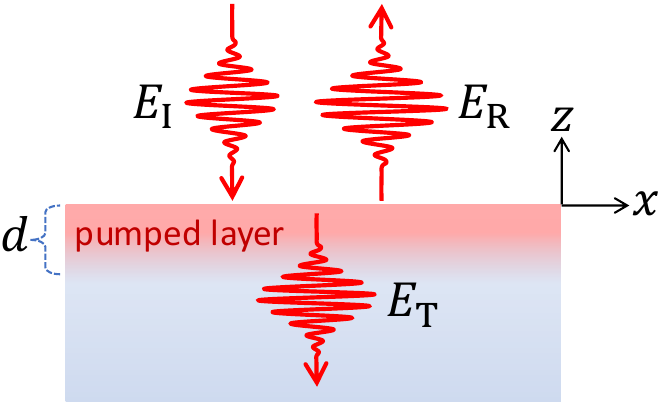}
	\caption{
		Schematic illustration of the reflection problem on the nonequilibrium sample.	
	}  \label{fig:sample_reflection}
\end{figure}

In the reflection problem, the sample lies in the region $z<0$ and the sample-vacuum interface is the x-y plane at $z=0$, as shown in \fig{fig:sample_reflection}.
The total EM field can be written as $\mathbf{E}=\mathbf{E}_{\mathrm{I}}+\mathbf{E}_{\mathrm{R}}$ at $z>0$ and $\mathbf{E}=\mathbf{E}_{\mathrm{T}}$ at $z<0$.
For simplicity, we consider a normally incident Gaussian light pulse $\mathbf{E}_{\mathrm{I}}\left(z,\, t\right)=E_0 e^{-(t-t_0+z/c)^2/2\delta^2-i\omega_0 (t+z/c)}$ with central frequency $\omega_0$. 
The optical reflectivity in the nonequilibrium problem can be defined as the ratio of the EM energy of the reflected pulse to that of the incident pulse\red{,}
\begin{equation}
	R=\frac{\int dt\left|\mathbf{E}_{\mathrm{R}}\left(0,\, t\right)\right|^{2}}{\int dt\left|\mathbf{E}_{\mathrm{I}}\left(0,\,t\right)\right|^{2}}.
\end{equation} 

In terms of the electric field $\mathbf{E}$, 
the Maxwell's equation for the transverse EM wave inside the nonequilibrium media ($z<0$) \red{reads}
\begin{align}\label{eq:EM_media}
\Box \mathbf{E} =-4\pi \partial_t \mathbf{j},
\quad
\mathbf{j}=\int_{t^{\prime}<t}dt^{\prime}
 \sigma\left(z; t,t^{\prime}\right)
\mathbf{E}\left(t^{\prime}\right)
\,
\end{align}
where $\Box = \partial_t^2 -c^2\nabla^2$, $c$ is the vacuum speed of light, $\mathbf{j}$ is the total  current and $\sigma\left(z; t,t^{\prime}\right)$ is the nonequilibrium optical conductivity.
Note that the time derivative of the current equals
$	\partial_t \mathbf{j}= \sigma\left(z; t,t\right)\mathbf{E}\left(t\right)+\int_{t^{\prime}<t}dt^{\prime}\left[\partial_{t} \sigma\left(z; t,t^{\prime}\right)\right]\mathbf{E}\left(t^{\prime}\right)
\,$.
The Maxwell's equation outside of the media ($z>0$) reads 
$ 
\Box \mathbf{E}
=0
$.
The reflected electromagnetic field $\mathbf{E}_{\mathrm{R}}$ can be obtained from the Maxwell's equations and the familiar boundary conditions on the interface:
$
E_{\parallel}(z=0^+) =E_{\parallel}(z=0^-),\,
B_{\parallel}(z=0^+) =B_{\parallel}(z=0^-)
$
where $E_{\parallel}$ ($B_{\parallel}$) means the in-plane component of the electric (magnetic) field.
The reflectivity in \fig{fig:rf}(b) is obtained from numerically solving \equa{eq:EM_media} and the boundary condition for the case of normal incidence and	 with the optical conductivity $\sigma\left(z; t,t^{\prime}\right)=e^{z/d}\sigma_{\mathrm{s}}(t,t^{\prime})+(1-e^{z/d})  \Theta(t-t^{\prime})  n_{s}^0e^2/m$.

We now provide an understanding of why the enhanced reflectivity may exceed unity in the limit of infinite penetration depth of the pump meaning that  the whole sample is uniformly prepared as a nonequilibrium superconductor.
Assuming an exponential dependence of the superfluid density on time:  $n_{\mathrm{s}}(t) \approx n_0 e^{ \gamma t}$, we take a derivative of the first equation of \equa{eq:EM_media} and obtain
\begin{equation}\label{eq:EM_media_2}   
\left(\partial_{t} -\gamma \right)
 \Box \mathbf{E}
=-\partial_{t}\left[\omega_{\mathrm{ps}}^{2}\left(t\right)\mathbf{E}\right]
\end{equation}
where  $\omega_{\mathrm{ps}}(t)=\sqrt{4\pi n_s(t) e^2 /m}$ is the instantaneous plasma frequency. 

In the case of $ \gamma \ll \omega_0$ and $\delta = \infty$, the picture is that of a continuous EM wave incident on a sample with a slowly changing dielectric. 
One may expect the EM wave below the surface of the sample to have the same  frequency ($\omega_0$) as  the incident beam, but with an instantaneous  wave vector $k(t)$  set by \equa{eq:EM_media_2} and corrected by   $\gamma$: 
$\mathbf{E}(\mathbf{r},t)=\mathbf{E}_{\text{T}} e^{-i\omega_0 t} e^{-i k(t) z}$.
At $z \ll 1/k$ and to leading order in $\gamma$, \equa{eq:EM_media_2} yields 
$
c^2 k(t)^2=\omega_{0}^{2} - \omega_{\mathrm{ps}}^{2}\left(t\right)
-i2 \gamma \omega_{\mathrm{ps}}^{2}/\omega_{0}$.
So as to reduce to the equilibrium case, we choose 
\begin{equation}\label{eq:rok}
ck=
	\begin{cases}
		i\sqrt{\omega_{\mathrm{ps}}^{2}-\omega_{0}^{2}
			+i2 \gamma \omega_{\mathrm{ps}}^{2}/\omega_{0}}\,, 
		& \omega_0<\omega_{\mathrm{ps}}(t),
		\\
		\sqrt{\omega_{0}^{2}
		-\omega_{\mathrm{ps}}^{2}
		-i2 \gamma \omega_{\mathrm{ps}}^{2}/\omega_{0}}\,, 
		& \omega_0>\omega_{\mathrm{ps}}(t).
	\end{cases}
\end{equation}
For normal incidence, the boundary condition on the interface reduces to 
$
E(z=0^+) =E(z=0^-),\,
\partial_z E(z=0^+) =\partial_z E(z=0^-)
$
which yields the Fresnel reflectivity 
\begin{equation}\label{eq:fresnel_reflectivity} 
	R=
	\left|
	\frac{\omega_{0}-ck}{\omega_{0}+ck}
	\right|^{2}
\end{equation}
in the continuous wave limit.
Therefore, for $\omega_0<\omega_{\mathrm{ps}}(t)$, one has	  $\mathrm{Im}[k]>0$, meaning an evanescent wave decaying exponentially into the bulk.
Notably, one has $R>1$ because 
$
\mathrm{Re}[k]<0
$
 which is caused by supercurrent growth.  
For $\omega_0>\omega_{\mathrm{ps}}(t)$, one has $\mathrm{Re}[k]>0$  such that the reflectivity is smaller than $1$.

A `quick and dirty' derivation of \equa{eq:rok} exists
using the model 
$
n_{\mathrm{s}}(t) \approx n_{s0} +[n_s(0)-n_{s0}]e^{- \gamma_\Delta t}
$
for the growing superfluid density, so that
the optical conductivity from \equa{eq:mfoc} has the form $
\sigma(\omega)|_{t=0}=\frac{e^2}{m} 
\left(i\frac{ n_{s0}}{\omega}+i\frac{n_s(0)-n_{s0}}{\omega+i\gamma_\Delta}\right)
$ 
in the frequency domain.
Note that the negative weight of the Drude term results in a negative $\mathrm{Re}[\sigma]$, meaning that the system is a gain medium in certain frequency range, a property  forbidden in equilibrium.
Around time zero, the wave vector of the EM wave inside the sample is therefore $ck=\sqrt{\epsilon(\omega_0)}\omega_0
=
\sqrt{1+\frac{4\pi i}{\omega_0}\sigma(\omega_0) }\omega_0
\approx	
\sqrt{\omega_{0}^{2}
-\omega_{\mathrm{ps}}^{2}
-i\gamma_\Delta (\omega_{\text{p}}^2-\omega_{\mathrm{ps}}^{2})/\omega_{0}}
$
where $\omega_{\text{p}}^2=4\pi n_{s0}(t) e^2 /m$.
The $\gamma_\Delta$ terms agrees with that of \equa{eq:rok} apart for a factor of $2$, if one uses the relation $\gamma_\Delta (\omega_{\text{p}}^2-\omega_{\mathrm{ps}}^{2})=\gamma \omega_{\mathrm{ps}}^{2}$ to make the initial growth rates of the superfluid density consistent between the two models.


\section{Enhanced reflectivity: perturbative approach}
\label{appendix:perturbation}
The enhanced   reflectivity could also be understood from a perturbative result for the reflected EM wave  in the nonequilibrium parameter $\gamma$.
This is doable in the situation that before time $t=0$,  a steady beam of monochromatic probe light $\mathbf E_i = \mathbf E_{i0}e^{- i \omega_0(t+z/c)}$ is already incident on the sample, resulting in steady transmitted ($\mathbf{E}^{(0)}$) and reflected beams. Starting at time $t=0$, the superfluid density grows with the rate $\gamma$.
Using $\omega_{\text{ps}}(0)$ as the frequency unit, we define the dimensionless variables: $t^\prime =\omega_{\text{ps}}(0) t$, $\gamma^\prime = \gamma/\omega_{\text{ps}}(0)$, 
$z^\prime=z \omega_{\text{ps}}(0)/c$, 
$\omega_0^\prime=\omega_0/\omega_{\text{ps}}(0)$ and work in the case $|\lambda| \ll 1$.
Assuming an exponential dependence of the superfluid density on time:  $n_{\mathrm{s}}(t) \approx n_0 e^{ \gamma t}$,
Eq.~\eqref{eq:EM_media} for $E(t^\prime,z^\prime)$  is now simplified to
\begin{equation}
	(\partial_{t^\prime}^2 -\partial_{z^\prime}^2) \mathbf E({t^\prime}) = -e^{\gamma^\prime {t^\prime} }\left[\mathbf E({t^\prime})
	+\gamma^\prime \int_0^{t^\prime}  d\tau  \mathbf E(\tau)\right].
	\label{S85}
\end{equation}
Laplace transforming Eq.~\eqref{S85} to the imaginary frequency $s$, we arrive at
\begin{equation}
	(s^2 -\partial_{z^\prime}^2) \hat E(s)+\frac{s}{s-\gamma^\prime}\hat E(s-\gamma^\prime)=s\mathbf E(0)+\partial_{{t^\prime}} \mathbf E(0).
	\label{S86}
\end{equation}
In the Taylor series $\hat E(s)=\hat E^{(0)}(s)+\hat E^{(1)}(s)+O(\gamma^2)$ in $\gamma$, the zeroth order transmitted electric field is given by
$
\mathbf E^{(0)}(t^\prime,\, z^\prime)
=\mathbf E_{i0}  
T
e^{- i(k_t^\prime z^\prime+\omega_0^\prime {t^\prime})}
$
where
$
k_t^\prime =\sqrt{\omega_0^{\prime 2}-1}
$
and 
$T=\frac{2\omega_0^\prime}{\omega_0^\prime+k_t^\prime}$
is the transmission coefficient.
In the Laplace domain, 
$
\hat E^0(s, z^\prime) 
= \mathbf E_{i0}  T
\frac{1 }{s+i\omega_0} e^{-\mathrm i k_t^\prime z^\prime}
$
. 
Because $\gamma$ is turned on right after $t=0$, the initial conditions are $\mathbf E(0) = \mathbf E^0(0)$ and $\partial_{t^\prime} \mathbf E(0) = \partial_{t^\prime} \mathbf E^0(0)$.
Assuming  the derivative $\partial_s E(s)$ is well defined, the $O(\gamma^\prime)$ terms in \equa{S86}  read
\begin{align}\label{eqn:EM_laplace}
	& (s^2 -\partial_{z^\prime}^2+1) \hat E^{(1)}(s)
	+ \frac{\gamma^\prime}{s} \hat E^{(0)}(s)
	- \gamma^\prime \partial_s \hat E^{(0)}(s)	
	=0
\end{align}
from which we obtain the $O(\gamma^\prime)$ transmitted electric field
\begin{align}\label{eqn:E1_laplace}
	\hat E^{(1)}(s,z^\prime)
	=
	\frac{\gamma^\prime  s}{s^2+\omega_0^2}\left[\frac{1}{(s+ i \omega_0)^2}-\frac{1}{s^2}\right]
	\frac{-2i e^{-\mathrm i k_t^\prime z^\prime} }{\omega_0+k_t^\prime}   \mathbf E_{i0} .
\end{align}
Transforming it back to time domain  and making use of the boundary condition on the interface, we obtain the reflected electric field:
\begin{widetext}
	\begin{align}
		\mathbf E_{\text{R}}({t^\prime})|_{z = 0} &= \mathbf E({t^\prime})|_{z=0} - \mathbf E_i({t^\prime})|_{z=0}
		=\mathbf E_{i0}e^{- i{\omega_0^\prime}{t^\prime}}
		\left[
		\frac{{\omega_0^\prime}-k_t^\prime }{{\omega_0^\prime}+k_t^\prime}
		-
		i\frac{2\gamma^\prime }{{\omega_0^\prime}+k_t^\prime }\left(\frac{{t^\prime}^2}{4}+\frac{i {t^\prime}}{4{\omega_0^\prime}}
		-\frac{e^{i{\omega_0^\prime}{t^\prime}}}{{\omega_0^\prime}^2}+\frac{5+3e^{2 i{\omega_0^\prime}{t^\prime}}}{8{\omega_0^\prime}^2}\right)
		\right]
		\label{S95}
	\end{align}
\end{widetext}
where the second term is the $O(\gamma^\prime)$ correction.
For $\omega_0 < \omega_{\text{ps}}(0)$, the wave vector
$
k_t^\prime =i \sqrt{1-\omega_0^{\prime 2}}
$ 
is a positive imaginary number so that the first term in \equa{S95} gives $R=1$ in equilibrium. 
Considering their signs, adding the first two terms in the $O(\gamma^\prime)$ correction obviously increases its norm, resulting in a reflectivity exceeding unity.
However, note that as a leading order result, it only works when $\gamma^\prime \ll 1$ and  $\gamma^\prime  t^{\prime 2} \ll1$.

\section{Linear EM response from the TDGL equation}
\label{appendix:TDGL_linear}
We use the gauge of zero scalar potential ($\phi=0$) and the EM field is represented by a vector potential $\mathbf{A}(\mathbf{r}, t)$.
In this section, we justify that as one takes the limit $\mathbf{q}=0$ continuously using the transverse vector potential,
the gradient of phase ($\nabla\theta$) in linear response to it would always be zero.  

We discuss the zero frequency and nonzero frequency cases of the driving vector potential $\mathbf{A}(\mathbf{r}, t)=\mathbf{A}e^{i(\mathbf{q}\cdot \mathbf{r}-\omega t)}$ separately.
At nonzero frequency ($\omega \neq 0$) for a transverse vector potential, it is natural that the phase $\theta$ does not respond to it.
Even for a longitudinal vector potential (note that a dynamical vector potential means nonzero electric field), one still has $\nabla\theta=0$ as one takes the limit $q \rightarrow 0$.
This is a general property of superconductors in a uniform vector potential oscillating at nonzero frequency: one needs some spatial in-homogeneity of a dynamical driving field to induce a phase gradient, see, e.g., Eq.~23 in Ref.~\cite{Sun.2020_collective}.
Therefore, as we take the limit $q \rightarrow 0$ for a transverse field, the picture goes smoothly to the well defined limit of dynamical uniform field, where $\nabla \theta$ in response to it is strictly zero.
At strictly zero frequency ($\omega=0$), the smooth limit of $\mathbf{q}\rightarrow 0$ does not agree with the other order of limit $(q=0,\, \omega\rightarrow 0)$ where there can indeed be a $\nabla\theta$ that cancels $\mathbf{A}$.
However, as long as $\mathbf{q}\neq 0$, one has $\nabla \theta=0$ strictly true for a transverse field. 
As one takes the smooth limit of $\mathbf{q}\rightarrow 0$, one obtains the correct limit for magnetic response of superconductors, i.e., the London equation for Meissner effect.
Therefore, the gradient of phase ($\nabla\theta$) won't exist in the limit we take.

For example, this can be verified directly from the TDGL in \equa{eq:tdgl}.
To compute the linear response of the order parameter  to the vector potential $\mathbf{A}(\mathbf{r}, t)$, one may write the order parameter in terms of amplitude and phase fluctuations around the mean field value: 
$\psi=(\psi_m+\delta)e^{i\theta}$ where $\psi_m$ satisfies
$\alpha \psi_m - 2 \psi_m^3=0$ and has been taken to be real without loss of generality.
Note that the vector potential is a real field and its plane wave representation $\mathbf{A}(\mathbf{r}, t)=\mathbf{A}e^{i(\mathbf{q}\cdot \mathbf{r}-\omega t)}$ is really intended to represent $\mathbf{A}(\mathbf{r}, t)=\mathbf{A}\cos(\mathbf{q}\cdot \mathbf{r}-\omega t)$.
At order $O(\mathbf{A}, \delta, \theta)$, the real and imaginary parts of TDGL give the
equations of motion for the amplitude and phase fluctuations:
\begin{align}
	&\frac{1}{\gamma}\partial_t \delta = (-2\alpha + \xi_0^2 \nabla^2)\delta - 
	\xi_0^2  \psi_m \nabla^2 \theta
	, \\
	&\frac{1}{\gamma}\partial_t \theta = \xi_0^2 \nabla^2 \theta-\xi_0^2 \psi_m
	(\nabla \cdot \mathbf{A})
	\,.
\end{align}
For the driving vector potential with frequency-momentum $(\omega, \mathbf{q})$, the linear response solution is 
\begin{align}\label{eqn:solution}
	\theta= \frac{-\xi_0^2  \psi_m}{-i\omega/\gamma+\xi_0^2q^2}
	\mathbf{q} \cdot \mathbf{A},
	\quad
	\delta = \frac{-(\xi_0^2  \psi_m)^2 q^2}{-i\omega/\gamma+2\alpha+\xi_0^2q^2}
	\mathbf{q} \cdot \mathbf{A}
	\,.
\end{align}
It is explicit that for a transverse field, one has $\nabla \theta=0$ because  $\mathbf{q} \cdot \mathbf{A}=0$.
Even for a longitudinal field, one has $\nabla \theta \rightarrow 0$ as $\mathbf{q}\rightarrow 0$ at a nonzero frequency $\omega$.

	\bibliography{references/Superconductivity}

\end{document}